\documentclass[11pt]{article}
\usepackage{hyperref}
\pdfoutput=1

\begin{document}
\title{Collapse of Non-axisymmetric Cavities}
\author{Oscar R. Enriquez, Ivo R. Peters, Stephan Gekle, Laura Schmidt,\\ Michel Versluis , Devaraj van der Meer, and Detlef Lohse\\
\\\vspace{5pt} Physics of Fluids Group \\ University of Twente, The Netherlands}
\maketitle
\begin{abstract}
 A round disk with a harmonic disturbance impacts on a water surface and creates a non-axisymmetric cavity which collapses under the influence of hydrostatic pressure. We use disks deformed with mode m=2 to m=6. For all mode numbers we find clear evidence for a phase inversion of the cavity wall during the collapse. We present a fluid dynamics video showing high speed imaging of different modes, pointing out the characteristic features during collapse.
\end{abstract}

\href{http://ecommons.library.cornell.edu/bitstream/1813/13967/2/MPEG-1.mpg}{Web version} (9.4 MB)

\vspace{2mm}

\href{http://ecommons.library.cornell.edu/bitstream/1813/13967/3/MPEG-2.mpg}{Full quality version} (221 MB)

\end{document}